\documentclass[dvips]{article}
\usepackage{subfigure}
\usepackage{here}
\usepackage{multirow}
\usepackage{amsmath,amssymb,graphicx,overcite}

\title{Development of a Compact Resonance Ionization Mass Spectrometer for Trace Element Analysis of Potassium}

\author{Yoshihiro Iwata, Yoshizumi Inoue$^{1}$, and Makoto Minowa$^{2}$}

\date{\it \small Experimental Fast Reactor Department, Oarai Research and Development Center, \\
Japan Atomic Energy Agency, 4002 Narita, Oarai, Ibaraki 311-1393\\
$^{1}$International Center for Elementary Particle Physics (ICEPP), University of Tokyo, \\
7-3-1 Hongo, Bunkyo, Tokyo 113-0033\\
$^{2}$Department of Physics, School of Science, University of Tokyo, 7-3-1 Hongo, Bunkyo, \\
Tokyo 113-0033}

\begin{document}
\maketitle

\begin{abstract}

A compact resonance ionization mass spectrometer is developed using 
two laser diodes and a quadrupole mass spectrometer to perform trace element analysis of potassium. 
With the help of a narrow linewidth of the laser diode, the isotope shifts of $^{40}{\rm K}$ and 
$^{41}{\rm K}$ of the 405~nm line with respect to $^{39}{\rm K}$, corresponding to the transition 
of $4\,^2$S$_{1/2}\rightarrow$ $5\,^2$P$^\circ_{1/2}$, are measured to be $207\pm13$~MHz and 
$451\pm10$~MHz, respectively, by comparing them to the known 
hyperfine splitting widths of the $4\,^2{\rm S}_{1/2}$ ground state of each potassium isotope. 
The overall detection efficiency of an order of $\sim10^{-6}$ in our setup indicates 
the availability of RIMS to the analysis of the trace metal impurities on or in a certain material 
such as the contamination assessment of semiconductor wafers. 

\end{abstract}

\section{Introduction}
\label{Sec_Introduction}

Recent development in laser technology has brought remarkable progress in spectroscopy, analytical 
spectrochemistry and mass spectrometry \cite{Wendt}. As one of the promising techniques realized 
by improvement of laser performance, selective excitation and photoionization of 
a specific kind of atom can be achieved with resonant absorption of a photon (or photons) 
with corresponding wavelength. This resonance ionization spectroscopy (RIS) is combined with a 
conventional mass spectrometer such as a quadrupole mass spectrometer (QMS) or a time-of-flight 
mass spectrometer (TOF-MS) to form resonance ionization mass spectrometry (RIMS), which is now 
used in various fields including geochemistry \cite{RELAX} and nuclear engineering \cite{FFDL} 
for isotopic analysis of noble gas, and was once proposed for the detection of the double beta 
decay and the solar neutrino \cite{Double-beta,Solar}. 
Although RIMS has not relatively been a common analytical method yet, it is quite effective 
in terms of its insensitivity to unwanted atoms or molecules as well as the compact size of 
the apparatus compared to accelerator mass spectrometry (AMS), which is considered to be a method 
with the best detection limit but a lack of convenience \cite{ICP-MS-summary}. 

The main characteristic of RIMS, insensitivity to unwanted atoms or molecules, leads to 
the suppression of background noise level to obtain high sensitivity to a certain element. 
As for the detection of trace isotopes, the effect of the tail of adjacent mass peak in 
a mass spectrometer has much impact on the discrimination of a specific isotope, along with 
high sensitivity of the apparatus. 

Since the electronic state of an atom, dependent on each element, is also known 
to be slightly different among isotopes of the same element, this subtle difference can be 
discriminated by CW lasers with sufficiently narrow linewidth ($\lesssim 1\,{ \rm MHz}$) and 
cannot by pulsed lasers with linewidth of typically $\gtrsim 1\,{\rm GHz}$. 
Besides the general isotope discrimination by mass spectrometers like TOF-MS or QMS, 
additional selectivity among isotopes can therefore be obtained by using CW lasers in the 
resonance ionization part. 

In this paper, we report on measurement of isotope shifts of $^{40}{\rm K}$ and $^{41}{\rm K}$ 
of the 405 nm line with respect to $^{39}{\rm K}$ by our compact resonance ionization mass spectrometer 
consisting of two laser diodes and a QMS.  
The overall detection efficiency of the spectrometer is improved compared to our previous setup 
\cite{JJAP-before}, and we propose the application of RIMS to the analysis of the trace metal impurities 
on or in a certain material such as the contamination assessment of semiconductor wafers. 
Surface contamination, especially contamination of trace metals including potassium, can cause 
a critical issue as device features continue to shrink and gate dielectrics scale 
toward the atomic level \cite{Beebe}. 
During high-temperature processing steps, metals at the surface can diffuse 
into the silicon substrate and act as recombination centers, adding 
electronic states into the band gap of silicon and degrading minority 
carrier lifetimes. 
Surface metallic impurities can also adversely affect silicon oxidation 
rates and become incorporated in gate oxides, where they degrade gate oxide 
integrity by increasing leakage currents. 
Surface measurement of trace metal contamination is thus an important step 
in device manufacturing.

\section{Hyperfine Splitting and Isotope Shifts of Potassium}
\label{Sec_potassium}

\begin{figure}
\begin{center}
\begin{tabular}[b]{c}
\subfigure[Resonance ionization scheme of potassium]
{\includegraphics[height=5cm]{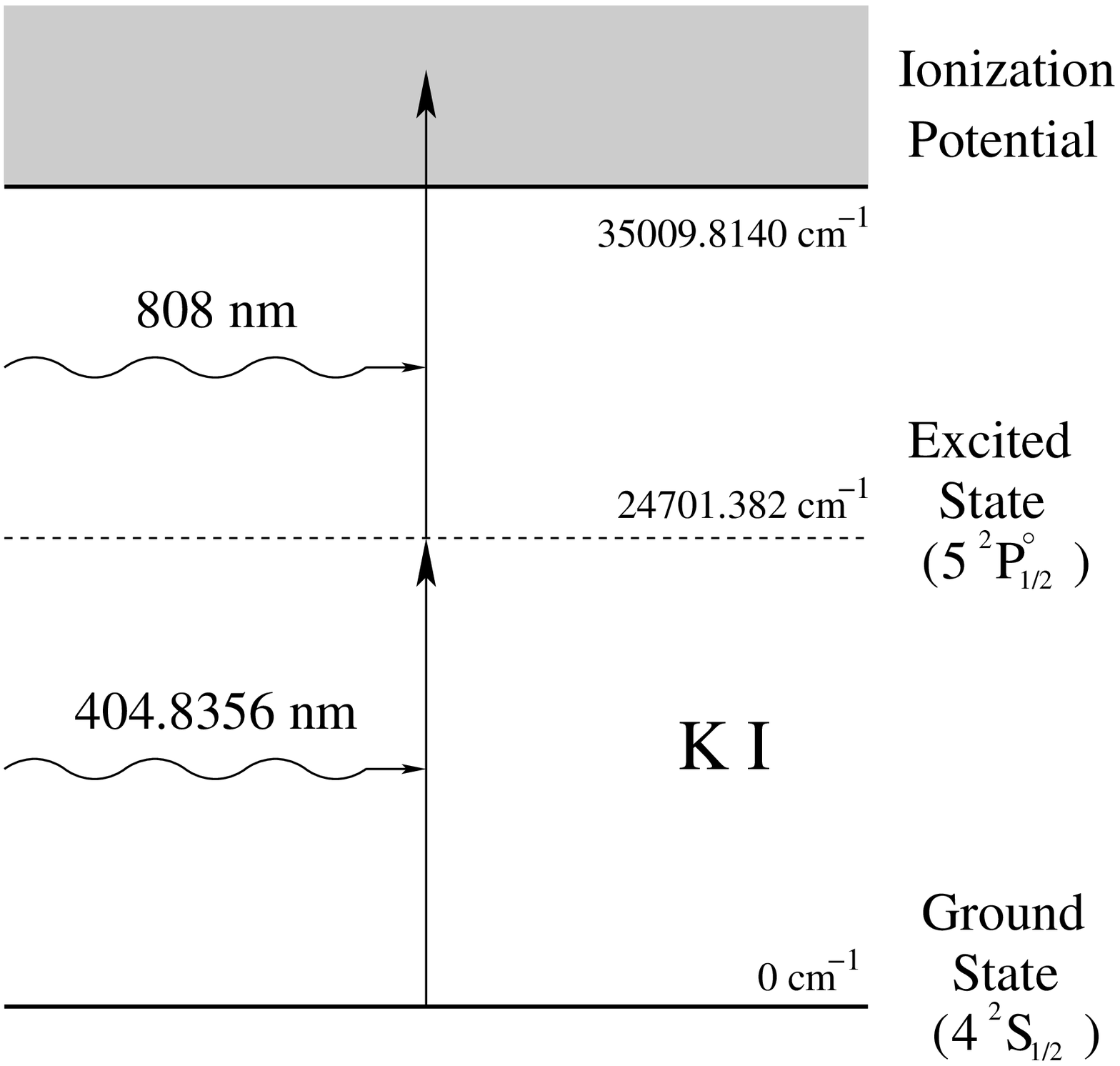}}
\hspace{0.8cm}
\subfigure[Hyperfine structures and isotope shifts of potassium isotopes relevant to this work]
{\includegraphics[height=5cm]{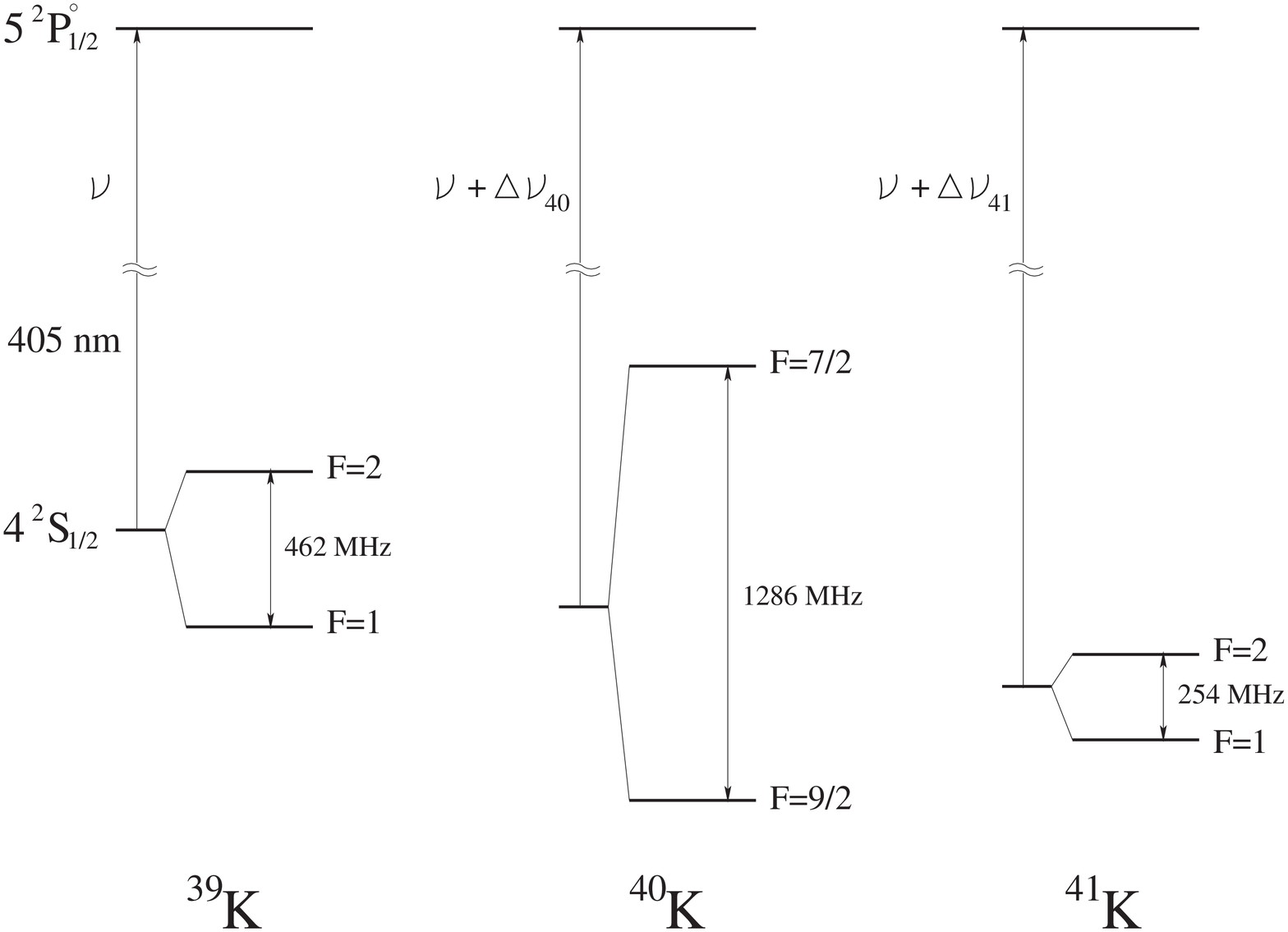}}
\end{tabular}
\end{center}
\caption{Electronic states of potassium isotopes.}
\label{RIMS-basic}
\end{figure}

Among various kinds of resonance ionization schemes to achieve photoionization of potassium 
\cite{RIS-K-ref}, the simplest two-photon RIS scheme, one-photon resonance excitation followed 
by one-photon ionization from the excited state performed by a single laser or two lasers, 
is employed for selective ionization of potassium, as shown in 
Fig.~\ref{RIMS-basic}-(a), from a cost-effectiveness point of view. 
The resonant wavelength of 404.8356~nm in vacuum is specific to potassium atoms, though 
it differs slightly among different isotopes because of the existence of hyperfine splitting 
and isotope shifts. 

\subsection{Hyperfine splitting}
\label{SubSec_HFS}

Hyperfine splitting (HFS) is caused by the interaction of the electron spin induced magnetic 
dipole moment with the magnetic moment of the atomic nucleus, whose energy width $\Delta E$, 
often divided by the Planck constant $h$ to be expressed in frequency unit, is given as follows
\cite{Diagnostics}: 
\begin{eqnarray}
\Delta E=\frac{1}{2}AC+B\frac{3C(C+1)-4I(I+1)J(J+1)}{8I(2I-1)J(2J-1)},
\label{HFS-eq}
\end{eqnarray}
where $I$ and $J$ are the nuclear spin and the total electronic angular momentum, respectively, and 
\begin{eqnarray}
C=F(F+1)-I(I+1)-J(J+1).
\end{eqnarray}
$F$ is the total angular momentum including nuclear spin and can take 
the following values:
\begin{eqnarray}
F=J+I,\,J+I-1,\cdots,|J-I|.
\end{eqnarray}
The HFS parameters $A$ and $B$ in Eq.~(\ref{HFS-eq}) are called the 
hyperfine structure interaction constants and correspond to the 
magnetic dipole and electric quadrupole constants, respectively. 
In a nonrelativistic treatment, $A$ and $B$ are written for 
the $n\ell$ electron state of alkali atoms to be \cite{Arimondo}:
\begin{eqnarray}
A&=&\left\{\begin{array}{ccc}
\displaystyle{-\frac{1}{h}\cdot\frac{\mu_0}{4\pi}\cdot\mu_{\rm B}^2\cdot
\frac{2\ell(\ell+1)}{J(J+1)}\cdot g_I\cdot
\langle r^{-3}\rangle_{n\ell}} && (\ell>0)\\
\displaystyle{-\frac{1}{h}\cdot\frac{16\pi}{3}\cdot\frac{\mu_0}{4\pi}\cdot
\mu_{\rm B}^2\cdot g_I\cdot\left|\Psi_s(0)\right|^2} && (\ell=0)
\end{array}\right.\\
B&=&\frac{1}{h}\cdot\frac{e^2}{4\pi\epsilon_0}\cdot\frac{2J-1}{2J+2}
\cdot\langle r^{-3}\rangle_{n\ell}\cdot Q,
\end{eqnarray}
where $n$ and $\ell$ are the principal quantum number and 
the orbital angular momentum quantum number, $\mu_0$ and $\mu_{\rm B}$ are 
the vacuum susceptibility and the Bohr magneton, $g_I$ is the nuclear 
$g$ factor, $\langle r^{-3}\rangle_{n\ell}$ is the average over the 
wave function of the electron $n\ell$ state, 
$\Psi_s(0)$ is the value of the Sch$\ddot{\rm o}$dinger wave function 
at the nucleus position and the scalar quantity $Q$ is conventionally 
taken as a measure of the nuclear quadrupole moment, respectively. 

The quadrupole term $B$ cannot be observed if $J=1/2$ as in the case of the $4\,^2$S$_{1/2}$ 
ground state of each potassium isotope. 
The energy splitting width is therefore given by 
$\Delta E=\frac{1}{2}(2I+1)A=$ $2A$, $\frac{9}{2}A$ and $2A$ for $^{39}$K, $^{40}$K and $^{41}$K, 
respectively. 
Figure~\ref{RIMS-basic}-(b) shows the hyperfine structure of the 
$4\,{^2{\rm S}_{1/2}}$ ground state of each potassium isotope \cite{Arimondo}. 
The $4\, ^2{\rm S}_{1/2}$ HFS widths of 462 MHz, 254 MHz and 1286 MHz for $^{39}{\rm K}$, 
$^{40}{\rm K}$ and $^{41}{\rm K}$, respectively, are known with high accuracy, 
which are also shown in this figure.

\subsection{Isotope shifts of $^{40}$K and $^{41}$K}
\label{SubSec_IS}

The isotope shift is a slight difference of the transition frequency
(wavelength) between the specific pair of the atomic states of two isotopes. 
This frequency shift $\Delta\nu$ is expressed as the sum of the normal mass shift (NMS), 
the specific mass shift (SMS) and the field shift (FS) as follows \cite{Behr}:
\begin{eqnarray}
\Delta\nu=\Delta\nu_{\rm NMS}+\Delta\nu_{\rm SMS}+\Delta\nu_{\rm FS}.
\end{eqnarray} 
The normal mass shift and the specific mass shift, collectively called the mass shift, 
are the reduced mass correction scaling with the transition frequency and the shift arising 
from the correlated motion of different pairs of atomic electrons, respectively. 
The field shift is produced by the different nuclear potentials coming from the 
different charge distributions. 
The value of the isotope shift of a specific isotope differs among different transitions. 

Isotope shifts of the 405 nm line of $^{40}{\rm K}$ and $^{41}{\rm K}$ with respect to 
$^{39}{\rm K}$ are denoted as $\Delta\nu_{40}$ and $\Delta\nu_{41}$, respectively, 
in Fig.~\ref{RIMS-basic}-(b). 
The measured value of $\Delta\nu_{41}$ is reported to be $456.1\pm0.8$~MHz using 
saturation spectroscopy \cite{Behr}. 
This method is often used in the measurement of isotope shifts or 
hyperfine structures because of its Doppler-free nature, but it seems rather difficult to 
measure the $^{40}{\rm K}$ signal with the isotope abundance of $10^{-4}$.

\section{Experimental Setup}
\label{Sec_Upgrade}

\begin{figure}
\begin{center}
\includegraphics[height=7cm]{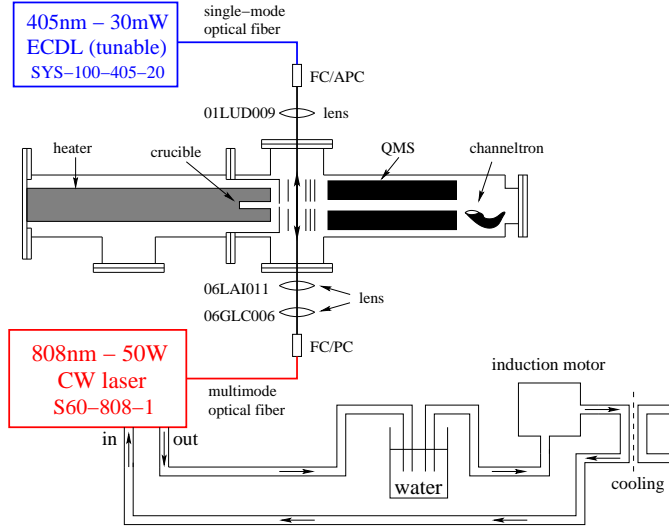}
\end{center}
\caption{Schematic view of the experimental setup in this work.}
\label{setup}
\end{figure}

The experimental setup in this study is described here. The basic structure has already been 
published \cite{JJAP-before}, and the laser system is newly upgraded mainly 
for the purpose of improving the photoionization efficiency as well as 
performing precise measurement of the $4\,^2$S$_{1/2}$ 
hyperfine structures and isotope shifts of the 405~nm line of potassium isotopes. 
A schematic view of the current experimental setup is shown in Fig.~\ref{setup}. 

Resonance ionization of potassium gas atoms obtained from the decomposition of K$_2$CO$_3$ sample 
in the electrothermally-heated graphite crucible is accomplished by 
two commercially available laser diodes. 
One is a wavelength-tunable external cavity diode laser (ECDL; Sacher Lasertechnik SYS-100-405-20) 
for the resonant transition of $4\,^2$S$_{1/2}\rightarrow$ $5\,^2$P$^\circ_{1/2}$. 
The wavelength is fine-tuned to $\lambda=404.8356$~nm in vacuum via the piezo actuator attached 
on the diffraction grating. 
An Elmos model AWG-50 waveform generator is newly prepared to scan the piezo voltage $V_{\rm piezo}$ 
over $\sim1$~V, corresponding to the laser frequency of $\sim2$ GHz. 
The laser output of up to $\sim$20~mW in the ionization region is provided through 
a single-mode fiber. 

The other is an Apollo Instruments model S60-808-1 water-cooled laser diode 
upgraded from an Amonics model ALD-808-3000-B-FC one to improve the photoionization efficiency.  
The laser output of up to $\sim$30~W in the ionization region is obtained through 
a multi-mode fiber. 

The generated potassium ions are extracted, mass analyzed by a Pfeiffer Vacuum model QMS200 QMS, 
and finally detected as an amplified ion current by the off-axis channeltron detector.

\section{Results and Discussion}
\label{Sec_Result}

\subsection{Measurements of $^{40}{\rm K}$ and $^{41}{\rm K}$ isotope shifts} 
\label{SubSec_Result_IS}

The isotope shifts of $^{40}{\rm K}$ and $^{41}{\rm K}$ of the 405 nm resonance excitation line 
with respect to $^{39}{\rm K}$, shown as $\Delta\nu_{40}$ and $\Delta\nu_{41}$ in 
Fig.~\ref{RIMS-basic}-(b), are measured by comparing them to the known hyperfine splitting 
widths of the $4\,^2{\rm S}_{1/2}$ ground state of each potassium isotope. 
Scanning the laser frequency over 2~GHz via the piezo voltage $V_{\rm piezo}$, 
each isotope shift is shown as the relative position of the weighted average frequency of 
two hyperfine splitting peaks of the targeted isotope to that of $^{39}{\rm K}$.  

\begin{figure}
\begin{center}
\begin{tabular}[b]{c}
\subfigure[$^{39}{\rm K}$ data]{\includegraphics[height=5cm]{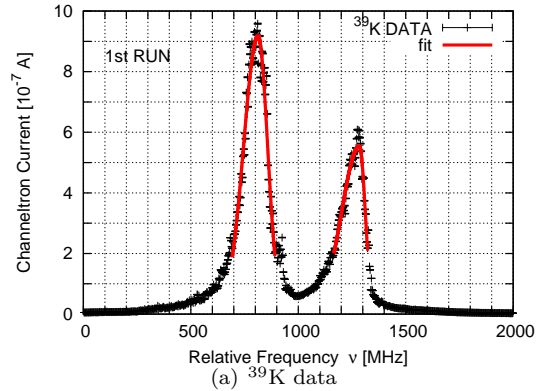}}\\
\subfigure[$^{40}{\rm K}$ data]{\includegraphics[height=5cm]{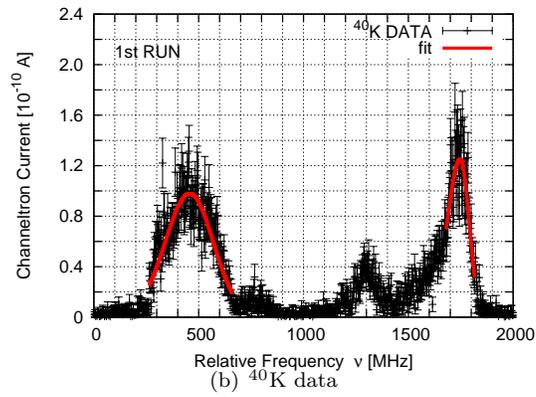}}\\
\subfigure[$^{41}{\rm K}$ data]{\includegraphics[height=5cm]{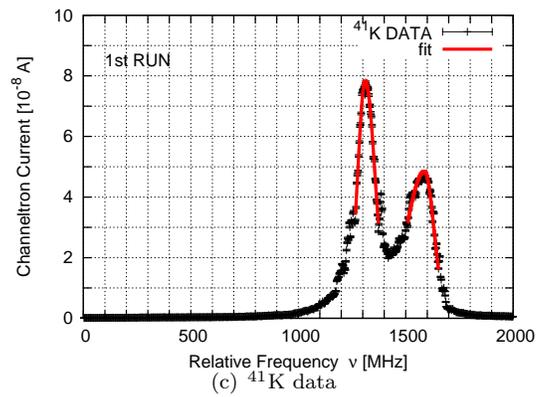}}\\
\end{tabular}
\end{center}
\caption{Measurement of the hyperfine structure of potassium 4s ground state for the 1st run with 
asymmetric Gaussian fitted curves.}
\label{HFS-result}
\end{figure}

An example of the experimental results is shown in Fig.~\ref{HFS-result} by scanning the piezo 
voltage in the range of $95.41\,{\rm V}\le V_{\rm piezo}\le96.26\,{\rm V}$. 
The crucible inner temperature was 1,020$\pm10$~K, 
the 405~nm laser power was $\sim20$~mW, the 808~nm laser power 
was $\sim2$~W and the applied channeltron voltage was 3~kV. 
The irrelevant peak seen at the relative frequency $\nu$ of around 1,300~MHz in the $^{40}{\rm K}$ 
spectrum (Fig.~\ref{HFS-result}-(b)) is due to the influence of the tail of the adjacent 
($^{39}{\rm K}$ or $^{41}{\rm K}$) mass peaks.
Contribution of a specific mass peak to the adjacent mass number is then seen to be about 10 ppm 
around the mass number of 40.

Fitted curves are also given in this figure, assuming that each absorption spectrum shows an 
asymmetric Gaussian profile around the resonant peak as expressed in Eq.~(\ref{asymmetric-eq}). 
This asymmetry is probably caused by the misalignment of the $\phi1$~mm orifice placed between 
the crucible and the ionization region. 
\begin{eqnarray}
I_{\rm ch}=\left\{
\begin{array}{ccc}
\displaystyle{I_0\exp\left[-\left(\frac{\nu-b}{c}\right)^2\right]} 
&& (\nu\le b)\\\\
\displaystyle{I_0\exp\left[-\left(\frac{\nu-b}{d}\right)^2\right]}
&& (\nu\ge b)
\end{array}\right.
\label{asymmetric-eq}
\end{eqnarray}
A total of 10 runs were performed under the same conditions. 
During the total measurement time of about an hour through 10 runs, the frequency drift was 
observed as slight differences of the relative frequency position among the obtained spectra. 
The horizontal scale in Fig.~\ref{HFS-result} is expressed in laser frequency unit 
which is transformed from raw $V_{\rm piezo}$ data using the experimentally obtained piezo 
response of $(2.39\pm0.05)\times10^3$~[MHz/V]. This linearity was obtained from the fitted relation 
between the piezo voltage difference $\Delta V_{\rm piezo}$ of the hyperfine splitting peaks 
and the known $4\, ^2{\rm S}_{1/2}$ HFS widths of 462 MHz, 254 MHz and 1286 MHz 
for $^{39}{\rm K}$, $^{40}{\rm K}$ and $^{41}{\rm K}$, respectively \cite{Arimondo}. 

The isotope shifts of $^{40}{\rm K}$ and $^{41}{\rm K}$ with respect to $^{39}{\rm K}$ are 
obtained as fitted averages of 10 experimental data to be $\Delta\nu_{40}=207\pm13$~MHz 
and $\Delta\nu_{41}=451\pm10$~MHz, the latter of which is consistent with the higher precision 
data by L. J. S. Halloran {\it et al.} using standard saturation spectroscopy \cite{Behr}. 
There have been so far no reports on $^{40}{\rm K}$ isotope shift of the 405~nm line. 
The quoted errors are statistical only. Additional systematic errors of $\pm$10\% were estimated 
by comparison of the obtained average over 10 data of 
(HFS width)$/\Delta V_{\rm piezo}$ among each HFS width of $^{39}{\rm K}$, $^{40}{\rm K}$ and 
$^{41}{\rm K}$ described above. 
They are mainly caused by the laser frequency fluctuation and nonlinearity of the piezo response. 
Application of a confocal Fabry-Perot interferometer frequency locking system instead of the piezo 
voltage to control the laser frequency with an accuracy of better than $\sim1$ MHz would help to 
perform more precise measurement \cite{CFI1,CFI2}. 

\subsection{RIMS overall detection efficiency}
\label{SubSec_Result_Efficiency}

The overall detection efficiency is defined as the ratio of the number $N_{\rm D}$ of 
potassium atoms detected by the QMS to that initially loaded in the crucible. 
Since it is hard to lock the laser frequency for a long time in our current setup, 
RIMS overall detection efficiency $E_{\rm RIMS}$ in this work was evaluated by the 
measurements of the ratio $R$ of the detection efficiency by RIMS to 
the detection efficiency by electron impact (EI) ionization. 
The absolute detection efficiency $E_{\rm EI}$ of EI ionization is estimated separately by switching 
the lasers off and filament for EI on. $E_{\rm RIMS}$ is then estimated to be 
$E_{\rm RIMS}=R\times E_{\rm EI}$. 
Estimation of the detection efficiency $E_{\rm EI}\sim8\times10^{-10}$ has already been 
measured in our previous work \cite{JJAP-before}. 

Measurement of the ratio $R$ of the detection efficiency by RIMS to 
the detection efficiency by EI ionization was performed 
using the upgraded laser system. 
The crucible heating current was set to 7.35~A in advance when the crucible inner temperature 
was 1,120$\pm10$~K. 
Resonance ionization signal could be observed at the 405~nm laser power 
output of $\sim15$ mW in the ionization region and the applied piezo 
voltage $V_{\rm piezo}$ of $\sim94$ V. 
The laser current of the upgraded 808~nm LD was set to 50~A to obtain the power output of 
$\sim30$~W in the ionization region. 
The channeltron voltage was set lower to 2~kV to avoid the saturation 
of channeltron currents observed at more than $10^{-5}$~A. 

\begin{figure}
\begin{center}
\begin{tabular}[b]{c}
\subfigure[Experimental result of the channeltron current of $^{39}\rm K^+$ atomic ions 
by RIMS and the channeltron current by electron impact (EI) ionization.]
{\includegraphics[height=4.5cm]{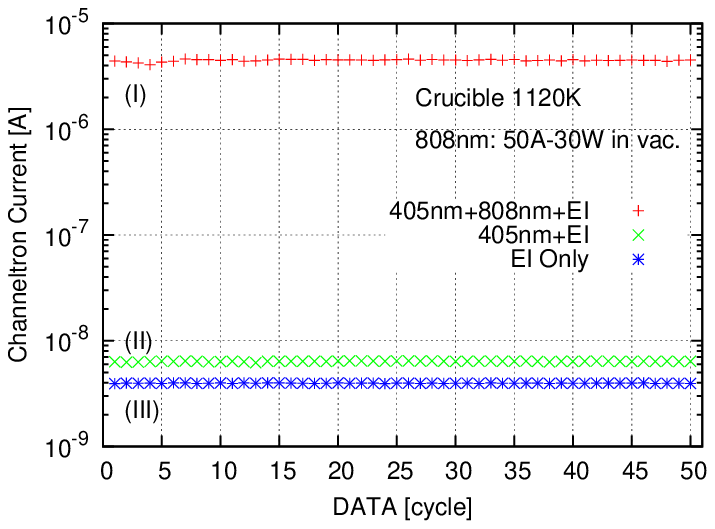}}
\hspace{0.8cm}
\subfigure[Ratios of the channeltron current without mass analysis to that 
with the mass resolution value set to 50.]
{\includegraphics[height=4.5cm]{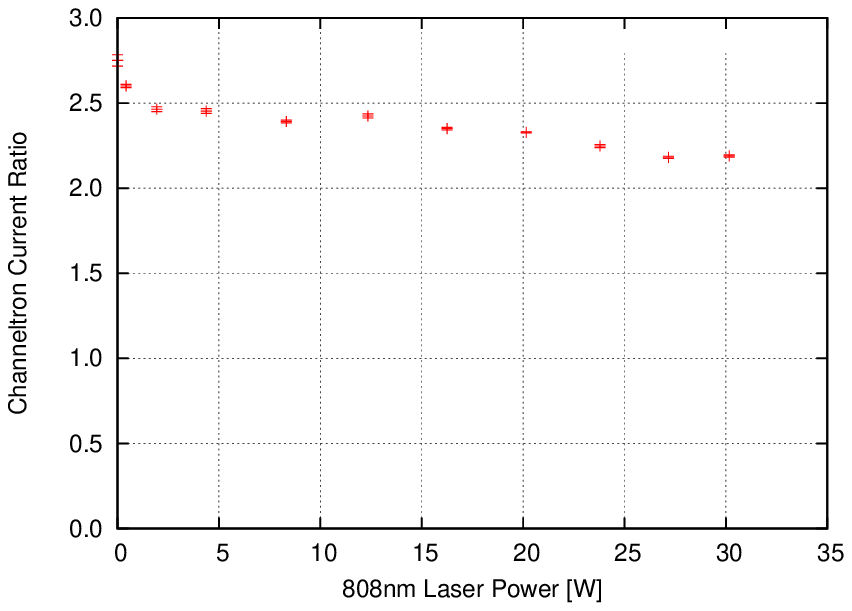}}
\end{tabular}
\end{center}
\caption{Experimental results regarding RIMS detection efficiency.}
\label{efficiency}
\end{figure}

Figure~\ref{efficiency}-(a) shows an experimental result of the channeltron 
current of $^{39}\rm K^+$ atomic ions by RIMS and the channeltron current by EI ionization. 
The measurement time is 20~ms in each cycle. 
Throughout the measurement, the emission current of the thermal electron 
for EI ionization was kept to 0.1~mA, in other words, the channeltron current contributed 
by EI ionization remained constant. 
Data taken when both lasers, 405~nm ECDL and 808~nm LD, were ON and 
only the 405~nm ECDL was ON are denoted as (I) and (II), respectively 
in the figure. (III) indicates the data for EI ionization as a reference. 
The ratio $R$ of the detection efficiency by RIMS to EI can be obtained 
as the ratio of the measured channeltron current of (I) with (III) subtracted to (III), giving 
$R\simeq4.5\times10^{-6}{\rm A}/4.0\times10^{-9}{\rm A}\sim1.1\times10^3$.
Combined with $E_{\rm EI}\sim8\times10^{-10}$, RIMS overall 
detection efficiency $E_{\rm RIMS}$ is estimated to be: 
$E_{\rm RIMS}=R\times E_{\rm EI}\sim 9\times10^{-7}$, improved by about 13 times with 
the upgraded 808~nm laser diode. 

$E_{\rm RIMS}$ can be divided into the efficiency $E_{\rm RIS}$ in the resonance ionization process and 
the transport efficiency $E_{\rm QMS}$ in the QMS: $E_{\rm RIMS}=E_{\rm RIS}\times E_{\rm QMS}$. 
Figure~\ref{efficiency}-(b) shows the obtained ratios of the channeltron current without 
mass analysis to that with it for various 808~nm laser power settings.  
The crucible heating current was set lower to 6.00~A when the crucible inner temperature was 
960$\pm10$~K and the channeltron voltage was set to 3~kV. The 405~nm laser power was unchanged. 
The transport efficiency $E_{\rm QMS}$ in the current setup, inverse of the obtained ratio shown in 
Fig.~\ref{efficiency}-(b), was measured to be a bit lower than 0.5, and was almost independent of 
the 808~nm laser power output up to 30~W. 
The efficiency in the RIS part is then estimated to be $E_{\rm RIS}\sim2\times10^{-6}$.

\section{Conclusion and Future Prospects}
\label{Sec_Conclusion}

We have developed a simple and convenient resonance ionization mass spectrometer for trace analysis of potassium. 
An upgraded 808~nm laser diode and a waveform generator are newly used for improvement of the 
photoionization efficiency and precise measurement of the isotope shifts of potassium isotopes, respectively. 

The obtained isotope shifts of $^{40}{\rm K}$ and $^{41}{\rm K}$ of the 405~nm line with 
respect to $^{39}{\rm K}$ are $\Delta\nu_{40}=207\pm13$~MHz and $\Delta\nu_{41}=451\pm10$~MHz, 
the latter of which is consistent with the data by L. J. S. Halloran {\it et al.} using standard 
saturation spectroscopy \cite{Behr}. 
There has been no reports on $^{40}{\rm K}$ isotope shift of this line to the authors' knowledge, 
probably because of the small isotope abundance of $^{40}{\rm K}$, 
and the lower resonance excitation cross section than that of other transition lines 
such as the 767~nm $4\,{^2{\rm S}_{1/2}}\,\rightarrow$\, $4\,{^2{\rm P}^\circ_{3/2}}$ and 
770~nm $4\,{^2{\rm S}_{1/2}}\,\rightarrow$\, $4\,{^2{\rm P}^\circ_{1/2}}$ lines. 

The obtained overall detection efficiency $E_{\rm RIMS}$ in our setup was 
$E_{\rm RIMS}=9\times10^{-7}$, in which the efficiency for resonance ionization was estimated to be 
$E_{\rm RIS}\sim2\times10^{-6}$. 
In the case of trace potassium analysis without isotope discrimination, 
mass analysis by a QMS is not indispensable and rather reduces the detection efficiency. 
Therefore, the estimated 3$\sigma$ detection limit $n_{\rm DL}$ is obtained to be 
\begin{eqnarray}
n_{\rm DL} = \frac{3\sqrt{\frac{I_{\rm BG}}{eG}\times T_{\rm meas.}}}{E_{\rm RIS}} \sim 
4\times10^8\hspace{0.3cm}{\rm [atoms]},
\end{eqnarray}
where $e$ is the elementary charge and $G\sim3.3\times10^5$ is the amplification gain of the channeltron.
The background channeltron current $I_{\rm BG}$ is here conservatively overestimated to be $\sim10^{-12}$~A, 
and $T_{\rm meas.}$ is taken to be $\sim1$ hour as a typical measurement time. 
The expected detection limit shows the availability of RIMS to the commercial use like the analysis of 
the trace metal impurities on or in a certain material such as the contamination assessment of 
semiconductor wafers. 
For the other impurities, there might be available ECDL's with suitable wavelengths. 
In case that there are no ECDL's with matched wavelengths, wavelength-variable dye lasers or 
Optical Parametric Oscillation (OPO) lasers can be used instead at a higher cost. 
The technology of the contamination assessment is also useful for the development 
of niobium (Nb) superconducting cavities for the International Linear Collider (ILC) project \cite{supercond}.

\section*{Acknowledgments}
This research is supported by a Grant-in-Aid for Exploratory Research from 
the Japan Society for Promotion of Science, by the Research Center for the 
Early Universe, School of Science, University of Tokyo, and also by the 
Global COE Program ``the Physical Sciences Frontier'' of the Ministry of 
Education, Culture, Sports, Science and Technology, Japan.

\end{document}